\documentclass[12pt]{article}
\usepackage{graphicx}

\newcommand\pubnumber{NuPhys2018-Drakopoulou}
\newcommand\pubdate{\today}

\def\napoli{School of Physics and Astronomy\\
University of Edinburgh, EH9 3FD, Scotland, United Kingdom \& \\
School of Physics and Astronomy\\ 
Queen Mary University of London, E1 4NS, London, United Kingdom}

\def\Title#1{\begin{center} {\Large #1 } \end{center}}
\def\Author#1{\begin{center}{ \sc #1} \end{center}}
\def\Address#1{\begin{center}{ \it #1} \end{center}}

\newcommand\pubblock{\rightline{\begin{tabular}{l} \pubnumber\\
         \pubdate  \end{tabular}}}
\newenvironment{Abstract}{\begin{quotation}  }{\end{quotation}}
\newenvironment{Presented}{\begin{quotation} \begin{center} 
             PRESENTED AT \end{center}\bigskip 
      \begin{center}\begin{large}}{\end{large}\end{center} \end{quotation}}

\def\beq{\begin{equation}}
\def\eeq#1{\label{#1}\end{equation}}
\def\eeqn{\end{equation}}

\def\beqa{\begin{eqnarray}}
\def\eeqa#1{\label{#1}\end{eqnarray}}
\def\eeqan{\end{eqnarray}}

\let\bar=\overbar

\def\Dslash{\not{\hbox{\kern-4pt $D$}}}
\def\dslash{\not{\hbox{\kern-2pt $\del$}}}

\def\msb{{\bar{\ssstyle M \kern -1pt S}}}

\begin{document}
\begin{titlepage}
\pubblock

\vfill
\Title{ANNIE Phase I: Neutron Background Measurements}
\vfill
\Author{ Evangelia Drakopoulou \& Benjamin Richards on behalf of the ANNIE Collaboration }
\Address{\napoli}
\vfill
\begin{Abstract}
The Accelerator Neutrino Neutron Interaction Experiment (ANNIE) is a 26-ton Gd-doped water Cherenkov detector installed in the Booster Neutrino Beam (BNB) at Fermilab. The primary physics goal of ANNIE is to study the multiplicity of final state neutrons from neutrino-nucleus interactions in water. Identifying and counting final state neutrons provides a new experimental handle to study systematic uncertainties related to the neutrino energy reconstruction in oscillation experiments. To achieve that goal ANNIE will make the first use of the novel Large Area Picosecond PhotoDetectors (LAPPDs). In Phase I, ANNIE characterised the beam-correlated neutron backgrounds in the detector and confirmed that they are sufficiently low for the Phase II physics measurements, anticipated to begin in 2019. In these proceesings the methodology and the results of Phase I will be discussed.
\end{Abstract}
\vfill
\begin{Presented}
NuPhys2018, Prospects in Neutrino Physics \\
Cavendish Conference Centre, London, UK, December 19--21, 2018
\end{Presented}
\vfill
\end{titlepage}
\def\thefootnote{\fnsymbol{footnote}}
\setcounter{footnote}{0}

\section{The ANNIE Experiment}

The ANNIE detector consists of a Gd-doped water detector deployed on the Booster Neutrino Beam at Fermilab \cite{BNB}. The beam is about 93\% pure $\nu_{\mu}$ (when running in neutrino mode) and has a spectrum that peaks at about 700 MeV. To reject entering backgrounds produced in the upstream rock a Front Muon Veto is used. An external Muon Range Detector downstream from the neutrino target  is utilised to range out and fit the energy and direction of muons from neutrino interactions in the water tank. The experiment was designed to proceed in two stages: a partially-instrumented test-beam run using only photomultiplier tubes (PMTs) and pure water (Phase I) for the purpose of measuring critical neutron backgrounds to the experiment \cite{ANNIE} and a physics run with a fully-instrumented Gd-doped water detector (Phase II). 

\begin{figure}[htb]
\centering
\includegraphics[height=3.in]{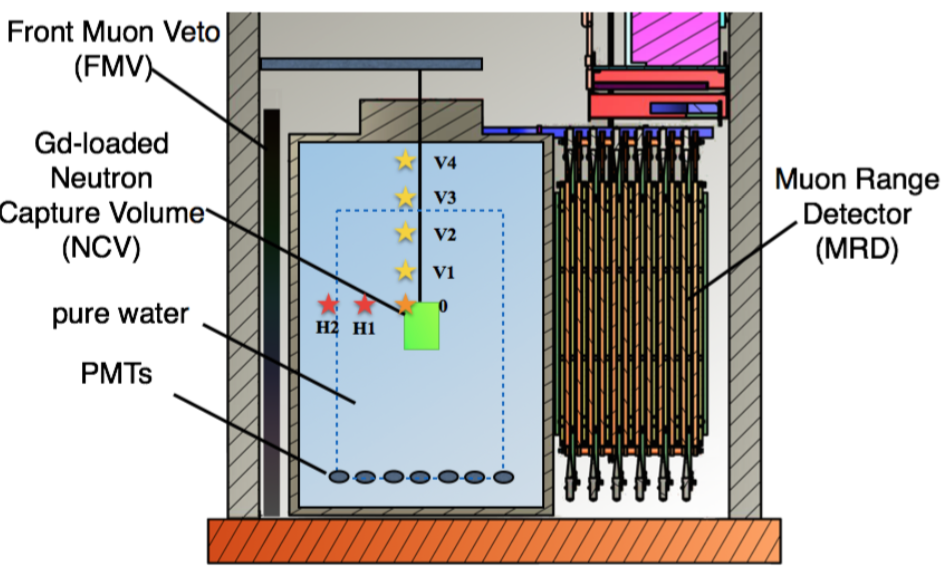}
\caption{A concept drawing of the Phase-I ANNIE detector system, showing the positions of the upper right corner of the NCV used in this analysis. The boundaries of the future Phase-II active volume are shown in the dashed line.}
\label{fig:BKGD}
\end{figure}

Phase I ended in late 2017 and the analysis of the data collected is now completed. The goal of Phase I was to measure and understand beam-induced neutron backgrounds. There are two potential background neutron sources in ANNIE: i) Sky-shine neutrons from the beam dump and ii) Dirt neutrons from the rock. 
Sky-shine neutrons are secondary neutrons produced in the beam dump that leak into the atmosphere and enter the detector after undergoing multiple scattering. Dirt neutrons are neutrons that arise from beam neutrino interactions occurring in the dirt and rock upstream of the experimental hall.  The ANNIE Phase I detector is shown in Fig. ~\ref{fig:BKGD}. For this phase, the tank was filled with 26 tons of ultra-pure deionised water. Inside the water volume, at the base of the tank, there is an array of 58 upward-facing 8" Hamamatsu R5912 PMTs. For the neutron background measurements two 3-inch PMTs in a 50 cm $\times$ 50 cm movable acrylic vessel with 25 gallons of Gd-loaded (0.25\% w/w) liquid scintillator are used. Thermal neutrons captured on Gd produce a cascade of photons with a total energy of around 8 MeV, which is visible as a bright flash of light in the scintillator. The neutron capture volume (NCV) was wrapped in reflective white plastic and completely enclosed in black plastic to optically isolate it from the rest of the tank. The detector neutron response was calibrated using $^{252}$Cf source. Then, the NCV was moved to different positions within the water volume to characterise the variation with position of the background neutron flux. 

\section{Results of ANNIE Phase I}
\begin{figure}[htb]
\centering
\includegraphics[height=3.5in]{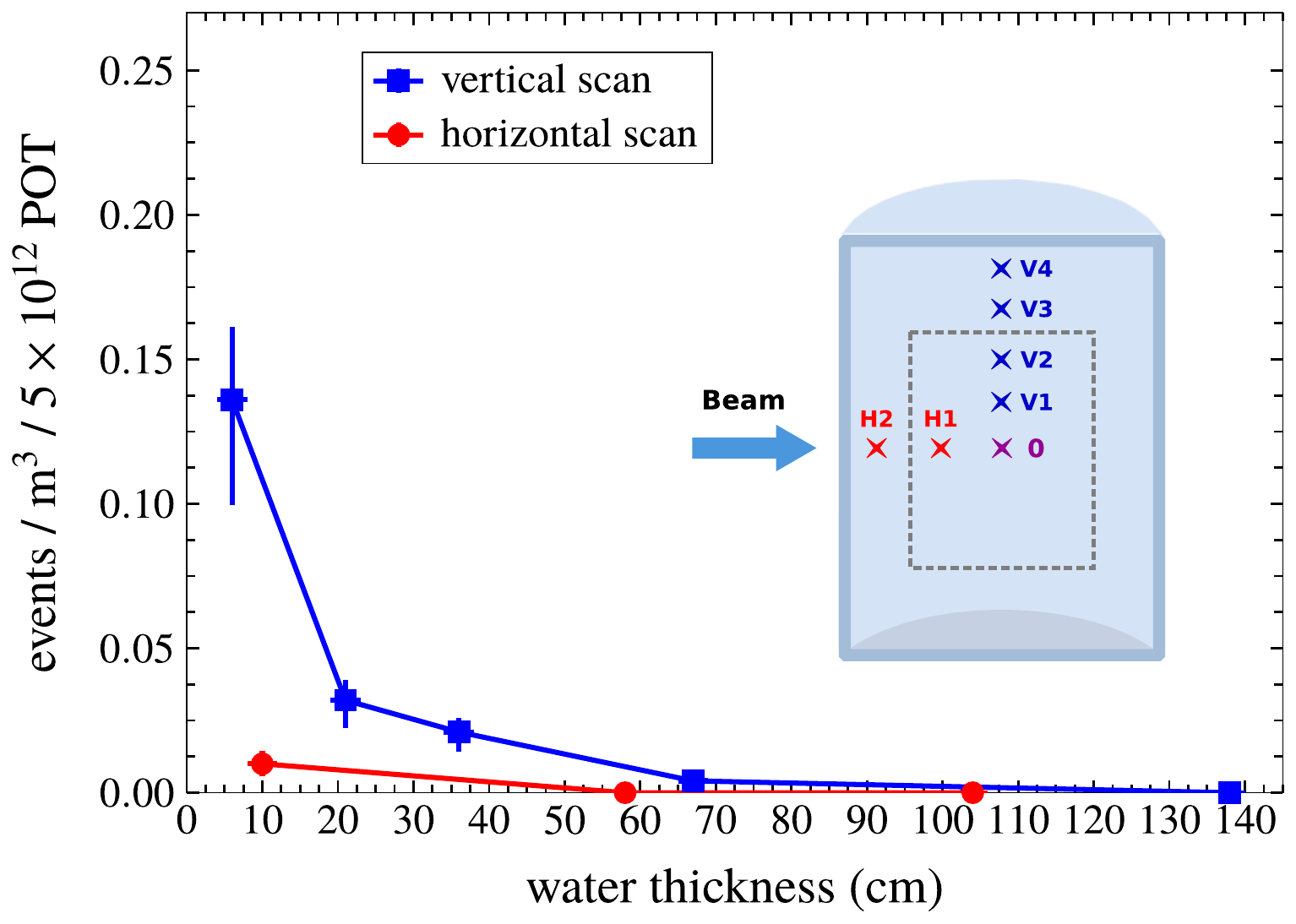}
\caption{Beam-correlated neutron candidate event rates measured during ANNIE Phase I. The inset diagram shows the NCV positions included in the red and blue datasets. The dashed line indicates which NCV positions are contained within the active region of ANNIE Phase-II. Position 0 (the centre of the tank) is shown in purple to indicate that it is included in both the red and blue data. For the blue dataset, the "water thickness" is the depth of the water above the top of the NCV. For the red dataset, it is the smallest distance between the side of the tube forming the NCV vessel and the beam side of the tank. The error bars shown in the plot include both statistical and systematic contributions.}
\label{fig:BKGDres}
\end{figure}

The background neutron flux at different locations in the tank was measured. The goal of ANNIE Phase-I was to determine the upper bounds on any beam induced neutron backgrounds that would inflate the count of signal neutrons in Phase-II, as well as understanding the sources of those backgrounds and how effectively they can be mitigated by shielding and isolating the Phase II active volume. Fig. ~\ref{fig:BKGDres} shows the beam-correlated neutron candidate event rates measured during ANNIE Phase I for different positions within the detector volume. We observe that the background event rate at position V4 (the top centre of the tank) is much larger than at all other NCV positions, including the most upstream location. This is consistent with the backgrounds being dominated by sky-shine neutrons rather than dirt neutrons. The rapid drop-off of these backgrounds with depth is consistent with the energy spectrum of these background neutrons being soft. A modest overburden of water, optically isolated from the main volume of the tank, will thus suffice to shield the Phase-II detector from the majority of background neutrons. The dashed line on the inset of this figure shows which NCV positions are located within the active detection volume of the ANNIE Phase-II detector.  The highest beam-induced background neutron rate within this active volume was measured at position V2, at 0.021 neutrons per m$^3$ per spill. This rate continued to drop with depth until position 0 where it is consistent with zero within errors.

The ANNIE Phase II detector is expected to see fewer than one interaction per spill at the BNB fluxes. Since neutrino interactions and background neutrons are statistically independent, the per-spill neutron rate can be thought of as the probability of detecting a background neutron following a signal neutrino interaction in Phase-II.

\section{ANNIE Phase II}
In ANNIE Phase II the detector will be fully instrumented with approximately 130 PMTs and at least 5 LAPPDs. The main physics goal of ANNIE in Phase II is to study the multiplicity of final state neutrons from neutrino-nucleus interactions in water. These measurements will improve our understanding of the many-body dynamics of neutrino-nucleus interactions and will allow to reduce the systematic uncertainties of the neutrino energy reconstruction in oscillation experiments and the signal-background separation for neutrino experiments. Efficient detection of neutrons in ANNIE will be made possible by searching for a delayed signal from their capture on Gd dissolved in water. Gd nuclei have high neutron capture cross-sections and produce 8 MeV gammas in $\sim$ 30 $\mu$s after the initial interaction, which provide a detectable signal in water Cherenkov detectors. ANNIE Phase II will be the first experiment to use fast-timing and position-precise LAPPDs \cite{LAPPD}, which have been produced by Incom Inc. and delivered to ANNIE, to perform these measurements. To realise the physics goals for Phase II, the ANNIE collaboration has developed several reconstruction techniques using the arrival time and position of the Cherenkov photons in the detector PMTs and LAPPDs. A maximum-likelihood fit is used to reconstruct the neutrino interaction vertex and direction. Deep Learning techniques are used to reconstruct the track length in the detector and Machine Learning algorithms are used for the muon and neutrino energy reconstruction~\cite{ANNIEReco}.


\section{Conclusions}
In Phase I, ANNIE demonstrated sufficiently low neutron backgrounds for the physics goals of Phase II. The key technological component of Phase II, LAPPDs, are now being produced by Incom Inc and have been delivered to ANNIE. The Phase II detector upgrade is ongoing and ANNIE Phase II physics data taking is foreseen in late 2019.


\end{document}